\documentclass[pra,twocolumn,tightenlines,nofootinbib]{revtex4-1}
\usepackage{bm,dcolumn,amsmath,graphicx,amsfonts,amssymb}

\usepackage{hyperref} 
\usepackage{xcolor}
\usepackage{comment}
\definecolor{cite}{rgb}{0.,0.,0.9}   
\hypersetup{colorlinks,linkcolor={cite},citecolor={cite},urlcolor={cite}}

\renewcommand{\v}[1]{\ensuremath{\boldsymbol{#1}}}		
\newcommand{\braket}[1]{\ensuremath{\langle #1\rangle}}	

\def\d{\ensuremath{{\rm d}}}

\newcommand{\smallspace}{\rule{0pt}{2.5ex}}

\usepackage[normalem]{ulem}
\usepackage{xspace}

\newcommand{\old}[1]{{\color{red}\sout{#1}}} %
\definecolor{newc}{rgb}{0.,0.6,0.4}

\newcommand{\note}[1]{{\color{orange}{${}^*$({#1})}}} 

\renewcommand{\note}[1]{}
\renewcommand{\old}[1]{}

\newcommand{\A}{\ensuremath{\mathcal{A}}} 

\begin{document} 

\title{The Bohr-Weisskopf effect: from hydrogenlike-ion experiments to heavy-atom calculations of the hyperfine structure}

\author{B.\ M.\ Roberts}\email[]{b.roberts@uq.edu.au}
\author{P.\ G.\ Ranclaud}
\author{J.\ S.\ M.\ Ginges}\email[]{j.ginges@uq.edu.au}
\affiliation{School of Mathematics and Physics, The University of Queensland, Brisbane QLD 4072, Australia}
\date{\today}

\begin{abstract}\noindent
In this paper we study the influence of electron screening on the Bohr-Weisskopf (BW) effect in many-electron atoms. The BW effect gives the finite-nucleus magnetization contribution to the hyperfine structure. Relativistic atomic many-body calculations are performed for $s$ and $p_{1/2}$ states of several systems of interest for studies of atomic parity violation and time-reversal-violating electric dipole moments --  Rb, Cs, Fr, Ba$^+$, Ra$^+$, and Tl. For $s$ states, electron screening effects are small, and the relative BW correction for hydrogenlike ions and neutral atoms is approximately the same. We relate the ground-state BW effect in H-like ions, which may be cleanly extracted from experiments, to the BW effect in $s$ and $p_{1/2}$ states of neutral and near neutral atoms through an electronic screening factor. This allows the BW effect extracted from measurements with H-like ions to be used, with screening factors, in atomic calculations without recourse to modelled nuclear structure input. 
It opens the way for unprecedented accuracy in accounting for the BW effect in heavy atoms. 
The efficacy of this approach is demonstrated using available experimental data for H-like and neutral $^{203}$Tl and $^{205}$Tl. 
\end{abstract} 

\maketitle


\section{Introduction}

Precision studies of the hyperfine structure in heavy atoms and ions play an important role in atomic and nuclear physics. 
They allow for stringent tests of quantum electrodynamics in strong electromagnetic fields~\cite{Volotka2013,Ullmann2017,Skripnikov2018}, determination of nuclear magnetic moments~\cite{Persson2013,RobertsFr2020,Konovalova2020,Barzakh2020} and tests of nuclear structure models~\cite{Shabaev1997,Beiersdorfer2001,Tomaselli2002,Senkov2002,Grossman1999,Zhang2015,Karpeshin2015,Prosnyak2021,Roberts2021}, 
as well as tests of atomic structure theory needed for precision atomic searches for physics beyond the standard model~\cite{GingesRev2004,RobertsReview2015,AtomicReview2017}.

In precision hyperfine calculations, one must account for the finite distribution of the nuclear magnetic moment across the nucleus, which gives a contribution to the hyperfine structure known as the Bohr-Weisskopf (BW) effect~\cite{Bohr1950,*Bohr1951}. 
For heavy systems, it is not currently possible to obtain this information from nuclear structure calculations with sufficiently high accuracy for a number of applications, and other methods are sought to deduce or control it. 
A notable example is in tests of bound-state quantum electrodynamics (QED) in hydrogenlike ions: the size of the BW effect is comparable to the size of the QED radiative corrections, and a method to remove the BW contribution in a specially-constructed difference of the effects in H-like and Li-like ions is utilized~\cite{Shabaev2001}. 

The last few years have seen a resurgence of interest in the Bohr-Weisskopf effect 
and in improved modelling of the nuclear magnetization distribution. 
This includes its use in the determination of nuclear magnetic moments~\cite{RobertsFr2020,Konovalova2020,Barzakh2020}, 
in understanding the neutron distribution in nuclei~\cite{Grossman1999,Zhang2015}, and in reliable tests of atomic wavefunctions in the nuclear vicinity~\cite{Ginges2018,Skripnikov2020}.  
The high sensitivity of the hyperfine structure to modelling of the finite nuclear magnetization distribution for a number of systems of interest for precision atomic tests of the standard model has only recently come to light~\cite{Ginges2017}. In some cases this amounts to a difference in the hyperfine structure of several percent. Inaccurate modelling of the nuclear magnetization distribution and BW effect has significant ramifications for fundamental physics tests, and the ability to test these models is critically important to these areas of study~\cite{Ginges2017,Roberts2021}. 

Progress has been made recently in testing nuclear magnetization models using available experimental data for heavy atoms. Studies of the differential hyperfine anomaly -- the difference in nuclear-size effects for different isotopes of the same system -- support the validity of the nuclear single-particle model for many of the considered atoms 
~\cite{Grossman1999,Zhang2015,RobertsFr2020,Barzakh2020,Demidov2020,Prosnyak2021,Roberts2021}.
The adoption of this model in place of the widely-used uniformly-magnetized ball model represents an advancement in the treatment of nuclear structure effects in heavy atoms 
and leads to a significant shift in the hyperfine structure in some cases. 

While the differential hyperfine anomaly gives a much-needed window into the nuclear magnetization distribution, it tests the {\it difference} in the BW effect between isotopes, while a test for a {\it single} isotope would provide a more powerful probe. 
For many-electron atoms, the atomic theory uncertainty limits the direct extraction of the BW effect from comparison of calculated and measured values for the hyperfine structure. Remarkably, this was carried out successfully~\cite{Skripnikov2020} for an isotope of Ra$^+$ for the unusual case in which the BW effect is several times larger than the atomic theory uncertainty~\cite{Skripnikov2020,Ginges2017}, 
with a BW uncertainty approaching $10\%$ (taking the atomic theory uncertainty to be 0.5\%).
Direct extraction for high-lying levels of heavy atoms is another possibility, where the atomic theory uncertainty is expected to be significantly smaller than for the lower levels~\cite{Ginges2018,Grunefeld2019}.     
To overcome limitations due to modelling of the BW effect in heavy atoms, it has been proposed to replace the effect with a ratio of measured and calculated values of the hyperfine structure~\cite{Ginges2018} 
in a similar vein to the specific difference method used to test QED~\cite{Shabaev2001} (see also discussion of the methods in Ref.~\cite{Karpeshin2021}).

In the current work, we consider extraction of the Bohr-Weisskopf effect from hydrogenlike ions for use in many-electron atoms. Indeed, the cleanest way to probe the magnetization distribution is through measurements with H-like ions, due to the high precision of both measurements and theory.   
Note that while in muonic atoms, the size of the BW effect may be an overwhelming $\sim 100\%$ the size of the total hyperfine interval, uncertainties connected to the experiments and theory hinder the high-precision determination of the effect~\cite{Buttgenbach1984,Crespo1998}. 
Measurements have been performed for several H-like ions of interest for tests of QED, including $^{209}$Bi$^{82+}$~\cite{Klaft1994}, $^{207}$Pb$^{81+}$~\cite{Seelig1998}, and $^{203,205}$Tl$^{80+}$~\cite{Beiersdorfer2001}. The Bohr-Weisskopf effects for these systems have been extracted with uncertainties at the level of 1\% 
(see e.g. Ref.~\cite{Senkov2002}, 
and Refs.~\cite{RobertsFr2020,Prosnyak2021} with updated nuclear magnetic moment values for $^{209}$Bi~\cite{Skripnikov2018} and $^{207}$Pb~\cite{Fella2020}). 

In this work, we study the effects of electron screening on the BW effect and calculate screening factors that relate the ground-state BW-effect in H-like ions to the effect in $s$ and $p_{1/2}$ states of neutral and near neutral atoms for systems of interest for precision atomic searches for new physics. 
This approach is based on the same principle highlighted in a recent work~\cite{Skripnikov2020} on the theory of the BW effect in molecules -- that the BW effect in atoms (and molecules) is determined fundamentally by the BW matrix element for the 1s state of the H-like system. 
The uncertainty of the screening factors for $s$ states of many-electron atoms is negligibly small. 
We demonstrate the validity of the approach with Tl, 
for which there is both H-like and neutral-atom precision hyperfine data available. 
It is hoped that this work will stimulate new experiments with H-like ions for the considered systems.


\section{The Bohr-Weisskopf effect}

The relativistic electron Hamiltonian for the interaction with the nuclear magnetic dipole moment is
\begin{equation}
\label{eq:h-hfs}
h_{\rm hfs}
= \alpha\, {\v{\mu}\cdot (\v{n}\times \v{\alpha})} \, F(r) /r^{2} ,
\end{equation}
where $\v{\alpha}$ is a Dirac matrix,
$\v{n}=\v{r}/r$ is the radial unit vector,
$\v{\mu}=\mu \v{I}/I $ with $\v{I}$ the nuclear spin and $\mu$ the magnetic moment, $\alpha\approx1/137$ is the fine structure constant, and $F(r)$ describes the nuclear magnetization distribution [$F(r)=1$ for the pointlike case]. 
We use atomic units ($\hbar$\,=\,$|e|$\,=\,$m_e$\,=\,1, $c$\,=\,$1/\alpha$) unless stated otherwise.
Matrix elements of the operator (\ref{eq:h-hfs}) may be expressed as $\A\braket{\v{I}\cdot\v{J}}$, where 
$\v{J}$ is the total electron angular momentum, and $\A$ is the magnetic dipole hyperfine constant.

The contribution to the hyperfine structure arising from account of the finite nuclear magnetization distribution is known as the Bohr-Weisskopf effect~\cite{Bohr1950}. This is a sizeable effect for heavy nuclei and typically enters at the percent level. In this work we use two nuclear magnetization models -- the uniform distribution (``ball'' model) and a simple nuclear single-particle model. Until recently it has been standard practice in the heavy atom community to use the ball model, 
\begin{equation}\label{eq:Fball}
F_{\rm Ball}(r)= (r/r_m)^3 \qquad \text{for} ~~~  r<r_m \, ,
\end{equation}
and $F_{\rm Ball}(r)=1$ for $r>r_m$.
A value for the nuclear magnetic radius, $r_m$, is usually found from the root-mean-square (rms) charge radius $r_{\rm rms}$, 
$r_m = \sqrt{5/3}\, r_{\rm rms}$. 
It has been shown recently \cite{Ginges2017,Ginges2018,RobertsFr2020,Roberts2021} that for a number of systems of particular interest in studies of fundamental symmetries violations, 
the ball model leads to sizeable errors in the calculated hyperfine constants by as much as several percent.  
A model that has been shown~\cite{RobertsFr2020,Prosnyak2021,Roberts2021} to be substantially more accurate for such systems is the simple nuclear single particle model~\cite{Bohr1950,Bohr1951,LeBellac1963,Shabaev1994}, which may be included in many-electron atomic calculations in a straightforward way~\cite{Tupitsyn2002},    
\begin{equation}\label{eq:FSP}
F_{\rm SP}(r) = F_{\rm Ball}(r)\left[1+ \Delta F(I,L,r/r_m)\right]. 
\end{equation}
Expressions for the term $\Delta F$, which depends on the nuclear spin, configuration, and magnetic moment, may be found in Refs.~\cite{Tupitsyn2002,volotka08a}. 
The simplest version of the model~\cite{volotka08a} -- which we use in this work -- takes the nucleon wavefunction to be constant across the nucleus, and excludes nuclear spin-orbit effects. More sophisticated modelling, in the Woods-Saxon potential and with spin-orbit interaction included~\cite{Shabaev1997,Artemyev2001}, leads to relatively small corrections for $^{87}$Rb, $^{133}$Cs, $^{211}$Fr~\cite{Ginges2017}, and isotopes of Tl ~\cite{Prosnyak2021} (see also Refs.~\cite{Shabaev1997,Gustavsson2000}).

It is convenient to express the hyperfine constant in the following form (see, e.g., Ref.~\cite{Shabaev1994}),
\begin{equation}\label{eq:BW}
\A = \A_0(1+\epsilon)+\delta \A_{\rm QED} ,
\end{equation}
with the Bohr-Weisskopf effect given as a relative contribution $\epsilon$. 
Here, $\A_0$ corresponds to the hyperfine constant found with a pointlike nuclear magnetization distribution ($F$\,=\,1) and with a finite nuclear charge distribution. We model the latter using a Fermi distribution, with the rms charge radii from Ref.~\cite{Angeli2013} and the thickness parameter taken to be 2.3\,fm. 
It is particularly useful to parametrize the BW effect as a relative rather than an absolute correction, as for heavy alkali-metal-like atoms and ions $\epsilon$ is, to a high degree, independent of: (i) electron correlation effects beyond core polarization; (ii) principal quantum number; and (iii) ionization degree (for $s$ states), as we explore further below. All of this is a consequence of the short-range-nature of the BW effect; see, e.g., Ref.~\cite{Shabaev2001}, and discussion below.   
The QED contribution to the hyperfine structure, $\delta \A_{\rm QED}$, must also be considered, as it enters with comparable size to the Bohr-Weisskopf effect~\cite{Shabaev1997} (see also, e.g., Refs~\cite{Sunnergren1998,Senkov2002,Sapirstein2003b,sapirstein06a,volotka08a,Volotka2012}).

For hydrogenlike ions, the simple atomic structure allows for calculations with particularly high precision.
Accurate determination of the Bohr-Weisskopf effect from experiments is therefore possible,
\begin{equation}\label{eq:extract}
\A_{\rm Expt.}^{1s}  = \A_0^{1s} (1+\epsilon^{1s} ) + \delta\A_{\rm QED}^{1s} ,
\end{equation}
as long as the QED contribution $\delta\A_{\rm QED}^{1s}$
and the nuclear magnetic moment --  which enters both terms on the right-hand-side of Eq.~(\ref{eq:extract}) -- are known sufficiently well. 
The BW effect is known to $\sim 1\%$ or better from measurements with  $^{203,205}$Tl$^{80+}$~\cite{Beiersdorfer2001}, $^{207}$Pb$^{81+}$~\cite{Seelig1998}, and $^{209}$Bi$^{82+}$~\cite{Ullmann2017}.

\subsection{Electron screening}

For many-electron atoms, the situation is different, and direct extraction of the BW effect from comparison of theory with experiment is strongly limited by uncertainties in the atomic structure that enter $\A_0$.
We proceed by introducing an electron screening factor, 
\begin{equation}\label{eq:xscreen}
x_{\rm scr.} = \epsilon/\epsilon^\text{H-like}, 
\end{equation}
and express the hyperfine constant for many-electron systems as
\begin{equation}\label{eq:AwithX}
A = \A_0(1+x_{\rm scr.}\,\epsilon^{\rm H-like})+\delta \A_{\rm QED},
\end{equation}
where $\epsilon^{\rm H-like}$ is the BW effect for the $1s$ or $2p_{1/2}$ state of the H-like ion of the same nucleus. 
For states with $j>1/2$, the Bohr-Weisskopf effect is essentially zero in the hydrogenlike case, and only becomes non-zero for many-electron atoms due to core polarization effects.
Therefore, it is only meaningful to define such screening factors relative to $j=1/2$ states of hydrogenlike ions.   
The screening factors depend only very weakly on the nuclear model and on atomic many-body effects beyond core polarization, as we explore further below, and they may therefore be determined with high accuracy. 

Indeed, the possibility to accurately determine such electronic screening factors forms the basis of the specific difference approach for removal of the BW effect in tests of QED, which has been formulated for Li-like ions and implemented for Li-like Bi \cite{shabaev1998,Shabaev2001,Skripnikov2018}.

\begin{table*}
\caption{Electron screening factors $x_{\rm scr.}$ [Eq.~(\ref{eq:xscreen})] for the lowest $s_{1/2}$ ($p_{1/2}$) states of several atoms of interest with respect to the $1s_{1/2}$ ($2p_{1/2}$) states of their H-like counterparts. The ratio $\eta_{sp}$ [Eq.~(\ref{eq:etasp})] for H-like systems is presented in the last row. Dash means value is consistent with zero within errors.}
\label{tab:xsp}
\begin{ruledtabular}
\begin{tabular}{l d d d d d d }
 & \multicolumn{1}{c}{$^{87}$Rb}     & \multicolumn{1}{c}{$^{133}$Cs}        &\multicolumn{1}{c}{$^{135}$Ba$^+$}       & \multicolumn{1}{c}{$^{203,205}$Tl}        &\multicolumn{1}{c}{$^{211}$Fr}       & \multicolumn{1}{c}{$^{225}$Ra$^+$}       \\
\hline\smallspace
$x_{\rm scr.}(s_{1/2})$  & 1.022 & 1.047 & 1.048(3) & 1.088 & 1.104 & 1.103(2) \\
$x_{\rm scr.}(p_{1/2})$  & \text{---} 
& 0.61(4) & 0.72(3) & 1.40(4) & 0.997(3) & 1.021(5) \\
$\eta_{sp}$ & 21.3   & 9.07(5)  & 8.65(1) & 3.60 & 3.02(1) & 2.93
\end{tabular}
\end{ruledtabular}
\end{table*}

\begin{figure*}
\includegraphics[width=0.333\textwidth]{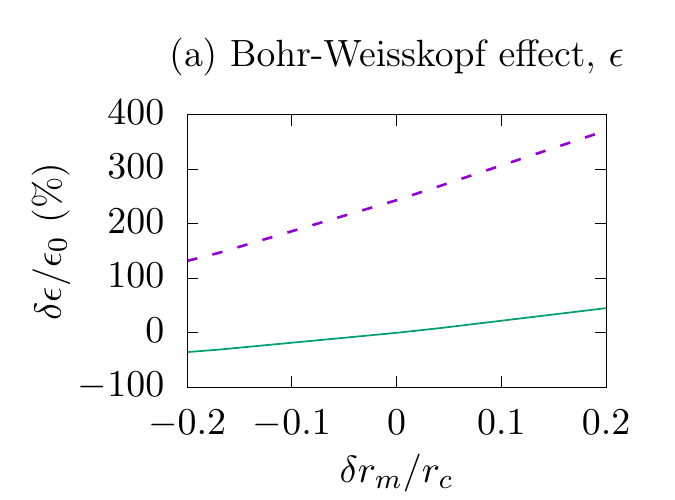}%
\includegraphics[width=0.333\textwidth]{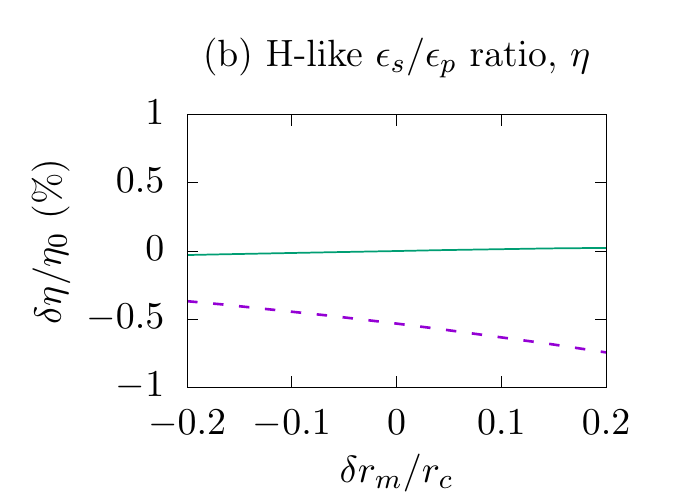}%
\includegraphics[width=0.333\textwidth]{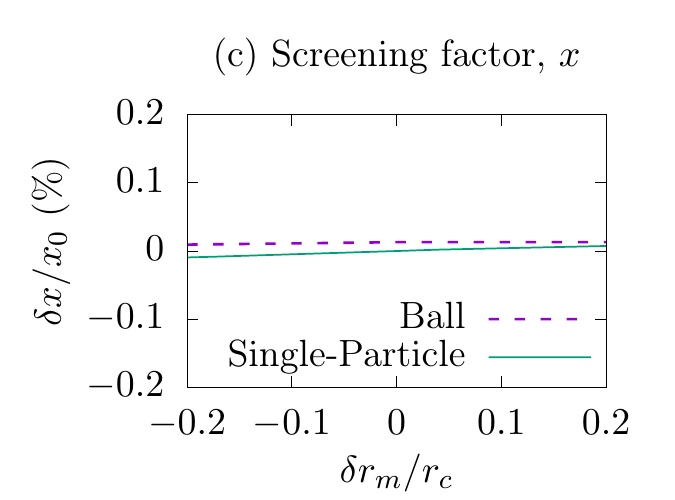}
\caption{Effect of relative changes in the nuclear magnetic radius for $^{133}$Cs on: 
(a) the $6s$ Bohr-Weisskopf effect, 
(b) the hydrogenlike $\epsilon(1s)/\epsilon(2p_{1/2})$ ratio,
and 
(c) the $6s$ screening factor.
The Bohr-Weisskopf effect depends strongly on the nuclear model and on the nuclear magnetic radius, while the screening factors and $\epsilon_s/\epsilon_p$ ratios do not.
The scale of the dependence is similar for other atoms.
Here, $\epsilon_0$, $x_0$, and $\eta_0$ correspond to nuclear single-particle model values 
found with magnetic radius $r_{m}=r_{c} = \sqrt{5/3}\,r_{\rm rms}$.
}
\label{fig:ex-plot}
\end{figure*}

To elucidate the behavior of the electron wavefunctions at small-distances and the BW effect,  consider the single-particle Dirac equation
$(H_D-\varepsilon)\phi = 0$, where
\begin{equation}
H_D = c \, \v{\alpha}\cdot\v{p}+(\beta-1)\,c^2 + V,
\end{equation}
with $\v{p}$ the electron momentum operator, $\beta$ a Dirac matrix, 
and $V$ the sum of nuclear and electronic potentials.
We express the single-particle electron orbitals as 
\begin{equation}\label{eq:DiracOrbital}
\phi_{n\kappa m} (\v{r}) = \frac{1}{r}
\begin{pmatrix}
f_{n\kappa}(r) \, \Omega_{\kappa m}(\v{n})\\
ig_{n\kappa}(r) \, \Omega_{-\kappa, m}(\v{n})
\end{pmatrix},
\end{equation}
where $f$ and $g$ are the large and small radial components of the orbital [normalized as $\int (f^2+g^2)\,\d r = 1$], $\Omega$ are two-component spherical spinors, $n$ is the principal quantum number, and
$\kappa=(l-j)(2j+1)$ is the Dirac quantum number (with $j$ and $l$ the total and orbital angular momentum quantum numbers, and $m=j_z$).
Then, for a spherically symmetric potential $V(r)$, the single-particle radial Dirac equation may be expressed for a given $\kappa$ as
\begin{equation}\label{eq:DiracMatrix}
\begin{bmatrix}
V(r)-\varepsilon				&  c(\kappa/r - \partial_r)\\
c(\kappa/r + \partial_r)	&  V(r)-\varepsilon-2c^2
\end{bmatrix}
\begin{bmatrix}
f_{n\kappa}\\
g_{n\kappa}
\end{bmatrix}=0.
\end{equation}

The Bohr-Weisskopf effect can be seen via Eqs.~(\ref{eq:h-hfs}) and (\ref{eq:BW}) 
to depend on the factor $F(r)-1$, which is non-zero only inside the nucleus:
\begin{equation}
\label{eq:epsilon}
\epsilon_{n\kappa} =
\frac{\int_0^{r_m} f_{n\kappa}(r)g_{n\kappa}(r) \,  [F(r)-1]/r^2 \, \d r}{\int f_{n\kappa}(r)g_{n\kappa}(r)  /r^2 \, \d r}.
\end{equation}
At small radial distances, $r$\,$\ll$\,$a_0/Z$, the electronic screening potential is negligible, and 
$V$\,$\approx$\,$-Z/r$.
It is seen from Eq.~(\ref{eq:DiracMatrix}) that for a given $\kappa$, the only state dependence comes from the energy $\varepsilon$. 
Since $|V|\gg|\varepsilon|$, the electron wavefunctions for each angular symmetry differ only by a multiplicative constant in this region, determined by the normalization of the wavefunction. This applies to states of different principal quantum number $n$ of the same atom, and is valid across all degrees of ionization of the atom, from neutral to H-like. 
In the context of the Bohr-Weisskopf effect, this behavior has been exploited in the formulation of the specific difference~\cite{Shabaev2001} and ratio~\cite{Ginges2018} methods; and the $n$-independence of the BW effect has been observed experimentally in neutral $^{85,87}$Rb \cite{PerezGalvan2007,PerezGalvan2008}.
%

Note furthermore that in the nuclear region the potential $|V|\gg mc^2$, and there is a symmetry between the $f_\kappa$ and $g_{\kappa'}$ components with $\kappa'=-\kappa$, as may be seen from Eq.~(\ref{eq:DiracMatrix}).   
Therefore, in this region $f_{s_{1/2}}\propto g_{p_{1/2}}$ and $f_{p_{1/2}}\propto g_{s_{1/2}}$, and the integrand in the numerator of Eq.~(\ref{eq:epsilon})
for $s$ and $p_{1/2}$ states is different only by a numerical factor. 
The above consideration means that the nuclear magnetization effects in atoms (and molecules) may be related fundamentally to the BW matrix element 
for the 1s state of the H-like system, a point that was highlighted in a recent work by Skripnikov~\cite{Skripnikov2020} in which the theory for molecules was set out and applied to RaF. 

Though the $s$-state Bohr-Weisskopf effect depends only weakly on correlation effects, the many-body effect known as core polarization (hyperfine correction to the states in the atomic core) gives an important contribution, particularly for states with $l>0$~\cite{Martensson-Pendrill1995}. 
We include this effect in the calculations via the relativistic time-dependent Hartree-Fock (TDHF) method, equivalent to the random phase approximation with exchange (RPA). 
Calculations are performed in the $V^{N-1}$ approximation ($N$ refers to the total number of electrons in the system), with the $N-1$ core electrons solved self-consistently in the average potential formed by the remaining core electrons.  Thallium is also treated using this approach, with the $6s^2$ subshell relegated to the core. Without core polarization, this is the relativistic Hartree-Fock (RHF) approximation and $V^{N-1}$ is the RHF potential. The valence electron wavefunctions and binding energies are found in this potential. Account of core polarization leads to a correction to the potential $\delta V$ that is first order in the external field (hyperfine interaction) and all-orders in the Coulomb
interaction. It is found by self-consistently solving the set of TDHF equations
\begin{equation}\label{eq:tdhf}
(H - \varepsilon_c)\delta\psi_c = -(h_{\rm hfs} +\delta V-\delta\varepsilon_c)\psi_c\, 
\end{equation}
for the core electrons. 
Here, $H$, $\psi_c$, and $\epsilon_c$ are the RHF Hamiltonian, core electron orbitals, and core electron binding energies, respectively, and $\delta\psi_c$ and $\delta\varepsilon_c$ are hyperfine-induced corrections for core orbitals and energies. The core polarization correction to the hyperfine structure is included by replacing the hyperfine operator $h_{\rm hfs}$ with $h_{\rm hfs}+\delta V$ in the valence electron matrix element.   
This well-known method has been described at length many times before, and we refer the reader to Ref.~\cite{Dzuba1987jpbRPA} and to recent works~\cite{Ginges2017,Grunefeld2019,RobertsFr2020,Roberts2021} for details about implementation for the hyperfine structure.

Our calculated electronic screening factors for the lowest $s$ and $p_{1/2}$ states of Rb, Cs, Fr, Tl, Ba$^+$, and Ra$^+$ are presented in Table~\ref{tab:xsp}. 
The BW effects for many-electron atoms were found from TDHF calculations for the hyperfine structure, the relative difference between results with the finite- and point-nucleus magnetization distributions yielding $\epsilon$; 
these values have been reported in our recent papers on the BW effect~\cite{Ginges2017,RobertsFr2020,Roberts2021}. 
The screening factors are found by taking the ratio of the BW result in atoms with that for the hydrogenlike counterpart. 
While the Bohr-Weisskopf effect depends strongly on the nuclear model and on the nuclear magnetic radius, the screening factors do not, which means they may be calculated with high accuracy without the need for a sophisticated nuclear model.
This is demonstrated in Fig.~\ref{fig:ex-plot}. 
To estimate the uncertainties in the screening factors, we include an error term equal to the largest difference between the values calculated using the single-particle and ball models, both with and without the inclusion of dominant correlation corrections beyond TDHF (the correlation potential; for details see Ref.~\cite{Dzuba1987jpbRPA}, and Refs.~\cite{Ginges2017,Grunefeld2019,RobertsFr2020} for recent implementations). 
On top of this, we include an error term equal to $5\%$ of the core polarization correction to $x_{\rm scr.}$.

\subsection{H-like s-p ratio}

Motivated by the symmetry between $f_{s/p}$ and $g_{p/s}$, and noting that (for hydrogenlike ions)
\begin{equation}
\A_0 (1+\epsilon) = {} \frac{2 g_I \, \mu_N \, \kappa}{j(j+1)} \int f_{n\kappa}(r)g_{n\kappa}(r) \,  F(r)/r^2 \, \d r,
\end{equation}
where $g_I = \mu/(\mu_N I)$ and $\mu_N$ is the nuclear magneton,
we construct the ratio
\begin{equation}\label{eq:etasp}
\eta_{sp}  = \epsilon^{1s} / \epsilon^{2p_{1/2}}
\end{equation}
for hydrogenlike ions. 
Due to the proportionality relations that hold in the nuclear region, $\eta$ is 
to a good approximation independent of principal quantum number and ionization degree.

Calculated values for $\eta_{sp}$ are given in Table~\ref{tab:xsp}. The results depend only to a small extent on the finite nuclear charge distribution. 
For example, for Cs$^{54+}$, the value for $\eta_{sp}$ (in the nuclear single-particle model) changes from 9.0 to 9.1
when the finite nuclear charge distribution is accounted for.
These ratios also depend only to a small extent on the nuclear magnetization model and effective magnetic radius.
For Cs$^{54+}$, the ratios differ by $\simeq$\,0.5\% between the ball and single-particle models; see Fig.~\ref{fig:ex-plot}.
For reasonable variations in the magnetic to charge radius,  
the resulting uncertainties are negligibly small. 
These have been taken into account in the uncertainty estimates in the results presented in Table~\ref{tab:xsp}. 
Our finding of the robust nature of the ratio to changes in nuclear structure is in agreement with calculations of the closely-related ratio $\epsilon ^{2s}/\epsilon ^{2p_{1/2}}$ that was the subject of a recent study~\cite{Konovalova2020}.  
In that work~\cite{Konovalova2020}, excellent agreement with the results of analytical calculations~\cite{Shabaev1994} was demonstrated, and the ratio was shown to be given by $(\epsilon ^{ns}/\epsilon ^{np_{1/2}})^{-1}=3/4\, Z^2\alpha^2$ to leading order. This is also correct to leading order for $1/\eta_{sp}$, and gives values that are accurate to about 10\%.  

If measurements are made of the hyperfine splitting of the ground-state of a hydrogenlike ion, the empirical value for the BW effect $\epsilon^{1s}$ may be accurately extracted [Eq.~(\ref{eq:extract})].
Then, the calculated ratio $\eta_{sp}$ may be used to determine the $p_{1/2}$ hydrogenlike BW effect.
These, together with the calculated screening factors $x_{\rm scr.}$, may be used to determine the BW effects in $s$ and $p_{1/2}$ states of many-electron atoms with high accuracy as follows: 
\begin{equation}\label{eq:AwithX}
\begin{split}
\A^s &= \A_0^s\left[1+x_{\rm scr.}^s \, \epsilon^{1s}  \right] +\delta \A_{\rm QED}^s, \\
\A^p  &= \A_0^p\left[1+ (x_{\rm scr.}^p/\eta_{sp}) \,  \epsilon^{1s}\right] +\delta \A_{\rm QED}^p .
\end{split}
\end{equation}
This removes entirely the nuclear structure uncertainty from atomic many-body calculations of the hyperfine structure. 

\section{Demonstration with thallium}

To demonstrate the method and its effectiveness, we consider two isotopes of thallium, $^{203}$Tl and $^{205}$Tl, that have been widely studied in the  context of the hyperfine structure and the BW effect; see the theoretical works~\cite{Shabaev1994,Martensson-Pendrill1995,Shabaev1997,Gustavsson2000,Tomaselli2002,Konovalova2017,Prosnyak2020,Prosnyak2021,Roberts2021}. 
Experimental data are available for both hydrogenlike ions and neutral atoms for these isotopes 
\cite{Lurio1956,Beiersdorfer2001,Chen2012}.


Empirical values for the BW effects in the $1s$ state of hydrogenlike $^{203}$Tl and $^{205}$Tl are presented in Table~\ref{tab:TlBW}. These were found from measured hyperfine splittings in these ions~\cite{Beiersdorfer2001}, along with the QED value calculated in Ref.~\cite{Shabaev1997} and the small Wichmann-Kroll magnetic loop correction from Ref.~\cite{Artemyev2001} amounting to $\delta \A_{\rm QED}^{1s} = -0.0116(1)\,(\mu/\mu_N)\,{\rm eV}$ (in excellent agreement with a value extrapolated from Ref.~\cite{Sunnergren1998}), together with our calculated values for $\A_0^{1s}$.   
Then, with our calculated ratio $\eta_{sp}$ presented in Table~\ref{tab:xsp}, we can infer with reasonably high accuracy the BW effect for the $2p_{1/2}$ state.
Finally, using the screening factors (Table~\ref{tab:xsp}), we can find the BW effect for the $6p_{1/2}$ and $7s_{1/2}$ states for neutral Tl.
These results are presented in Table~\ref{tab:TlBW}, along with calculated values for $\epsilon$. 
Note that for the considered Tl isotopes, the single-particle and ball models coincide, since they have spin and parity designations $I^\pi = 1/2^+$.

\begin{table}
\caption{BW effect, $\epsilon$ (\%), for $1s$ and $2p_{1/2}$ states of H-like Tl, and $6p_{1/2}$ and $7s$ states of neutral Tl.
Single-particle model (SP) values are shown alongside results found using the measured value for H-like $1s$.
}
\label{tab:TlBW}
\begin{ruledtabular}
\begin{tabular}{llllll}
   && $\epsilon$ (SP) & \multicolumn{3}{c}{$\epsilon$ (empirical)\tablenotemark[1]} \\
\cline{4-6}\smallspace
 &     &  Tl\tablenotemark[2] &$^{203}$Tl &  $^{205}$Tl & Method\\
\hline\smallspace
H-like&$1s_{1/2}$ &-1.95&-2.216(10)&-2.236(11) & Expt.\\
&$2p_{1/2}$ &-0.54&-0.616(3)&-0.621(3) & $\eta_{sp}$\\
\smallspace
Neutral&$6p_{1/2}$ &-0.78&-0.86(3)&-0.87(3)&  $\eta_{sp}$, $x_{\rm scr.}$\\
&$7s_{1/2}$ &-2.13&-2.41(1)&-2.43(1)&  $x_{\rm scr.}$
\end{tabular}
\tablenotetext[1]{Found using H-like $1s$ measurement~\cite{Beiersdorfer2001}, QED calculations \cite{Shabaev1997,Artemyev2001}, and atomic calculations ($\A_0$, $x$, and $\eta$) from this work.} 
\tablenotetext[2]{With $I^\pi=1/2^{+}$, the SP and ball models coincide for $^{203}$Tl and $^{205}$Tl, and the BW effect is nearly the same for these isotopes.}
\end{ruledtabular}
\end{table}

To test the empirically-deduced BW values we have obtained for neutral Tl, we consider the differential hyperfine anomaly~\cite{Persson1998},
\begin{equation}
{}^1\Delta^2 \equiv \frac{\A^{(1)}/g_I^{(1)}}{\A^{(2)}/g^{(2)}_I}-1
\approx \epsilon^{(1)} - \epsilon^{(2)} + \delta^{(1)}  - \delta^{(2)},
\end{equation}
for which there is accurate experimental data. The anomaly $^1\Delta^2$ is comprised of the differential finite nuclear magnetization (Bohr-Weisskopf) and differential finite nuclear charge (Breit-Rosenthal~\cite{Rosenthal1932}, $\delta$) effects.
It often happens that for nearby isotopes of the same atom, the  Breit-Rosenthal effects strongly cancel and the differential anomaly is dominated by the differential BW effect~\cite{Persson2013}. 
However, this is not always so, in particular when the nuclear spins of the isotopes are the same~\cite{Persson2013}, as is the case for the considered Tl isotopes.

In Table~\ref{tab:1D2-Tl}, we present the differential anomalies $^{203}\Delta ^{205}$ for H-like and neutral Tl.
The first column of results corresponds to calculations performed in the SP model. In the next column, the empirically-deduced results are presented, found from the Bohr-Weisskopf results derived from the hydrogenlike ion measurements shown in Table~\ref{tab:TlBW}.
The values include the contribution from the Breit-Rosenthal effect; indeed, this effect gives nearly the whole contribution to the
calculated single-particle values for the considered isotopes.  
Comparison with the measured values shows that the calculated SP results are too small by about a factor of two. 
On the other hand, the empirically-deduced results agree with the measured data.  
However, due to strong cancellations in the differential anomaly, the associated relative uncertainties are significantly larger than for the BW effect.

\begin{table}
\caption{Hyperfine anomaly $^{203}\Delta^{205}$ (\%) for the $1s$ and $2p_{1/2}$ states of H-like Tl, and the $6p_{1/2}$ and $7s$ states of neutral Tl.
The values `SP' and `via $1s$' are results of this work, the former found in the single-particle model, the latter found using the measured H-like $1s$ result from Table~\ref{tab:TlBW}; 
measured values are presented under `Experiment'.
We don't show the extracted value for $1s_{1/2}$, since this matches the experimental value by definition. 
}
\label{tab:1D2-Tl}
\begin{ruledtabular}
\begin{tabular}{lllllll}
 & & \multicolumn{1}{c}{SP} & \multicolumn{1}{c}{via $1s$} & \multicolumn{2}{c}{Experiment} \\
\hline\smallspace
H-like &$1s_{1/2}$ &0.014& --- &0.032(12) & \cite{Beiersdorfer2001} (2001)\\
&$2p_{1/2}$ &0.004 &0.008(4)& --- &\\
\smallspace
Neutral &$6p_{1/2}$ &0.005&0.012(6)&0.01036(3)&\cite{Lurio1956} (1956)\\
&	 				& 	& 					&0.0121(47)&\cite{Chen2012} (2012)\\
&$7s_{1/2}$ &0.014&0.034(14)&0.0294(81)&\cite{Chen2012} (2012)\\
\end{tabular}
\end{ruledtabular}
\end{table}

\subsection{Nuclear magnetic radius}

In the modelling of the nuclear magnetization distribution, the nuclear magnetic radius is often taken to correspond to that of the charge radius, in particular their root-mean-square radii are taken to be the same. 
However, this is a rough approximation, as typically the magnetization distribution does not come from the bulk of the nucleus, it arises largely from unpaired nucleons. 
Following a recent similar study by Prosnyak {\it et al.}~\cite{Prosnyak2020} (and earlier works on H-like $^{185,187}$Re~\cite{Beiersdorfer1998}, H-like $^{203,205}$Tl~\cite{Beiersdorfer2001}, and muonic $^{203,205}$Tl and $^{209}$Bi~\cite{Elizarov2005}), we introduce an effective magnetic radius, $r_{m}$, into Eqs.~(\ref{eq:Fball}) and (\ref{eq:FSP}) for $F(r)$, that is defined as the value required to reproduce the observed magnetic hyperfine anomalies.   
In general, this radius will depend on the model used for $F(r)$, and it also has some dependence on the modelling of the charge distribution.
Below, we will demonstrate this method using $^{203}$Tl and $^{205}$Tl. 

\begin{figure}
\includegraphics[width=0.48\textwidth]{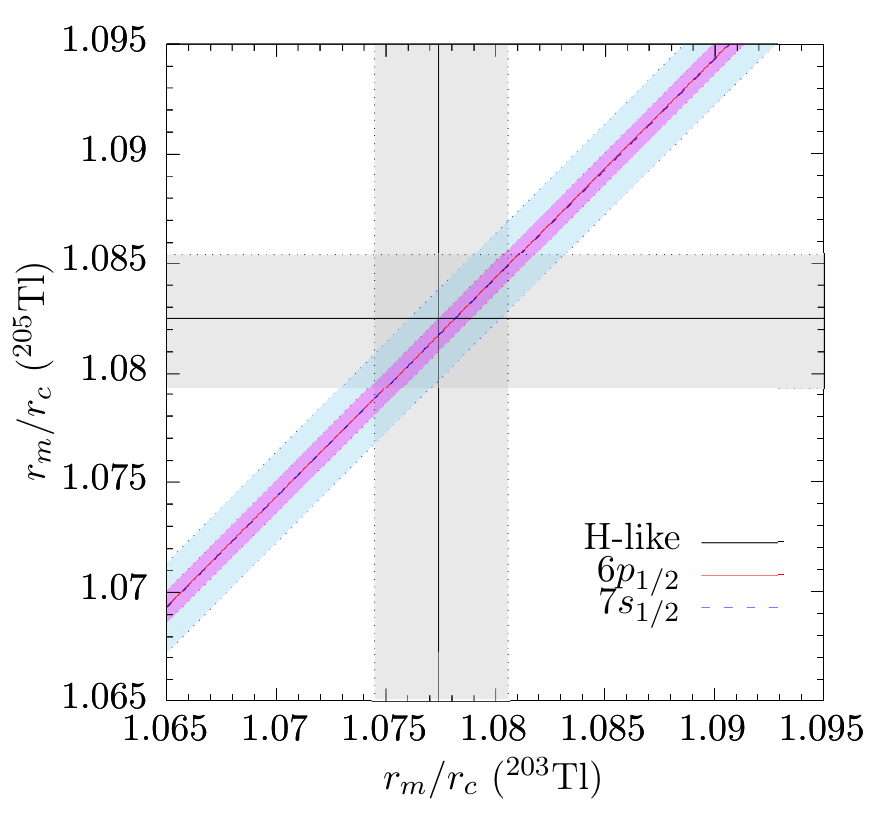}~~
\caption{Magnetization radius (in units of the 
reference radius 
$r_{c} = \sqrt{5/3}\,r_{\rm rms}$)
for $^{203,205}$Tl, as constrained by measurements of the H-like Bohr-Weisskopf effect and the neutral-atom differential hyperfine anomaly.
Solid black lines are from experimental BW effect for $^{203,205}$Tl$^{80+}$. 
Diagonal coloured lines relate $r_{m}$ for $^{203}$Tl and $^{205}$Tl deduced from experimental hyperfine anomaly for $6p_{1/2}$ (magenta) and $7s_{1/2}$ (blue) states.
Uncertainties are indicated by shaded regions.}
\label{fig:Tl-rmag}
\end{figure}

We proceed in two independent ways.
First, we find the value of $r_{m}$ that reproduces the observed $s$-state BW effect for the $^{203,205}$Tl hydrogenlike ions, accounting for the QED effects as above  [Eq.~(\ref{eq:extract})].
These results depend on values of individual magnetic moments, which we take from the compilation of Stone~\cite{Stone2005}.
This is shown by the horizontal/vertical black lines in Fig.~\ref{fig:Tl-rmag}.
We then consider the $6p_{1/2}$ and $7s_{1/2}$ differential hyperfine anomalies for the neutral atoms.
By fixing the magnetic radius for one isotope, we may deduce that required for the other isotope in order to reproduce the observed anomaly. 
The differential anomaly is particularly sensitive to the 
difference in the squared magnetic radii of the two isotopes~\cite{Martensson-Pendrill1995,Beiersdorfer2001,Prosnyak2020}, but cannot on its own lead to a determination of either magnetic radius, as can be seen in the diagonal lines in Fig.~\ref{fig:Tl-rmag}. 
Note that the differential anomaly depends on the ratio of the magnetic moments $\mu(^{205}{\rm Tl})/\mu(^{203}{\rm Tl})=1.00983613(6)$~\cite{Baker1963}, which is known to higher precision than the individual moments~\cite{Gustavsson1998}.
By combining the two sets of results from hydrogenlike and neutral atom data, the effective nuclear magnetic radii may be accurately determined.

The two lines from the neutral $6p$ and $7s$ anomalies in Fig.~\ref{fig:Tl-rmag} agree precisely, with smaller error bars for the $6p$ data.  
Note that this coincidence is not guaranteed, as the lines are calculated independently.
Indeed, at the Hartree-Fock level (i.e., without core polarization), the two lines do not coincide.
The shaded regions in the plot represent the ($1\sigma$) uncertainties.
For the hydrogenlike ions, the uncertainty is dominated by the QED calculations. 
For the neutral systems, the uncertainty comes both from the experimental determination of the anomalies and from the theoretical calculations connected mostly to the account of many-body effects. 
In particular, the former dominates the $7s$ result, and the latter dominates the $6p_{1/2}$. 
Our results give the following values for the effective magnetic radii: 
$r_m/r_c=1.077(3)$ for $^{203}$Tl, and $r_m/r_c=1.083(3)$ for $^{205}$Tl, and their ratio $r_m(^{205}{\rm Tl})/r_m(^{203}{\rm Tl})=1.0057(5)$. 
Our results for individual magnetic radii are smaller than those found in Ref.~\cite{Prosnyak2020}, while the inferred difference in the squared rms magnetic radii agrees with previous works \cite{Beiersdorfer2001,Prosnyak2020}; the deviation could be explained by the different modelling of the nuclear charge distribution.
As a final consistency check, we calculate the BW effect for $6p_{1/2}$ and $7s$ states of neutral Tl using the deduced effective magnetic radii. 
For $^{203}$Tl we find $\epsilon(6p) = -0.886\%$ and $\epsilon(7s) = -2.41\%$, 
and for $^{205}$Tl we find $\epsilon(6p) = -0.893\%$ and $\epsilon(7s) = -2.43\%$, in excellent agreement with the BW values determined from hydrogenlike measurements, shown in Table~\ref{tab:TlBW}.

\section{Conclusion}

We propose a joint theoretical and experimental scheme to accurately extract the Bohr-Weisskopf effect from H-like ions for use in calculations with heavy atoms. 
Using the presented method, the BW effect for many-electron atoms may be determined with high accuracy from a measurement of the ground-state hyperfine structure for the corresponding hydrogenlike ion.
This method allows one to remove nuclear uncertainties from calculations of the hyperfine structure, and to perform a clean and reliable test of atomic theory uncertainty in the nuclear region through hyperfine comparisons. This is important for precision atomic calculations needed in studies of atomic parity violation and time-reversal-violating electric dipole moments. 
At the same time, experimental determination of the Bohr-Weisskopf effect will aid in better understanding nuclear structure by providing a window for testing nuclear models, including the neutron distribution. We hope this work will stimulate new measurements with hydrogenlike ions, where experimental capability is rapidly expanding.

\acknowledgements
This work was supported by the Australian
Government through Australian Research Council (ARC)
DECRA Fellowship No. DE210101026 and ARC Future Fellowship
No. FT170100452.

\bibliography{hfs,other}

\end{document}